\documentclass{article}

\usepackage{graphicx}
\usepackage{oljour}
\usepackage{color}

\begin{document}


\journalname{zk}  

\title[de]{}

\title[en]{Reducing the positional  modulation of NbO$_6$-octahedra in Sr$_{x}$Ba$_{1-x}$Nb$_{2}$O$_{6}$
by increasing the Barium content: A single crystal neutron diffraction study at ambient temperature for 
$x$=0.61 and $x$=0.34}


\begin{author}
  \anumber{1}   
  \atitle{Dr.}    
  \firstname{J\"urg}  
  \surname{Schefer}    
  \vita{}      
  \institute{Laboratory for Neutron Scattering, ETH Zurich \& Paul Scherrer Insitute}   
  \street{Bldg.}    
  \number{WHGA-244}    
  \zip{5232 Villigen PSI}      
  \country{Switzerland}   
  \tel{+41-56-3104347}      
  \fax{+49-56-31023191}      
  \email{jurg.schefer@psi.ch}     
\end{author}

\begin{author}
  \anumber{2}   
  \atitle{Dr.}    
  \firstname{Dominik}  
  \surname{Schaniel}    
  \vita{}      
  \institute{I. Physikalisches Institut}   
  \street{Z\"ulpicher Strasse}    
  \number{77}    
  \zip{50937 K\"oln}      
  \country{Germany}   
  \tel{+49-221-4706355}      
  \fax{+49-221-4705162}      
  \email{dominik.schaniel@uni-koeln.de}     

\end{author}

\begin{author}
  \anumber{3}   
  \atitle{Dr.}    
  \firstname{Vaclav}  
  \surname{Pet\v{r}\'{\i}\v{c}ek}    
  \vita{}      
  \institute{Institute of Physics}   
  \street{Academy of Sciences of the Czech Republic}    
  \number{Na Slovance 2}    
  \zip{18221 Prague}      
  \country{Czech Republic}   
  \tel{+420-220-318598}      
  \fax{+420-233-343184}      
  \email{petricek@fzu.cz}     
\end{author}

\begin{author}
  \anumber{4}   
  \atitle{PD Dr.}    
  \firstname{Theo}  
  \surname{Woike}    
  \vita{}      
  \institute{I. Physikalisches Institut}   
  \street{Z\"ulpicher Strasse}    
  \number{77}    
  \zip{50937 K\"oln}      
  \country{Germany}   
  \tel{+49-221-4706355}      
  \fax{+49-221-4705162}      
  \email{Th.Woike@uni-koeln.de}     
\end{author}

\begin{author}

  \anumber{5}   
  \atitle{Dr.}    
  \firstname{Alain}  
  \surname{Cousson}    
  \vita{}      
  \institute{Laboratoire L\'eon Brillouin}   
  \street{CEA-CNRS}    
  \number{}    
  \zip{91191 Gif-sur-Yvette}      
  \country{France}   
  \tel{+33-169-086431}      
  \fax{+33-169-088261}      
  \email{cousson@llb.saclay.cea.fr}     
\end{author}

\begin{author}

\anumber{6}   
  \atitle{PD Dr.}    
  \firstname{Manfred}  
  \surname{W\"ohlecke}    
  \vita{}      
  \institute{Fachbereich Physik,University of Osnabr\"uck}   
  \street{Barbarastrasse}    
  \number{7}    
  \zip{49069 Osnabr\"uck}      
  \country{Germany}   
  \tel{+49-541-9692631}      
  \fax{+49-541-9693512}      
  \email{manfred.woehlecke@uos.de}     

\end{author}

\corresponding{jurg.schefer@psi.ch}

\abstract{We report on the influence of the Barium content on the modulation amplitude in  Sr$_{x}$Ba$_{1-x}$Nb$_{2}$O$_{6}$ compounds by comparing  Sr$_{0.61}$Ba$_{0.39}$Nb$_{2}$O$_{6}$ (SBN61) and Sr$_{0.34}$Ba$_{0.66}$Nb$_{2}$O$_{6}$ (SBN34).  Our single crystal neutron diffraction results demonstrate 
that the amplitude of the positional modulation of the NbO$_6$ octahedra is reduced with increasing barium content, indicating that the origin of the modulation is 
the partial occupation of the pentagonal channels by Sr and Ba atoms. By increasing the Sr content the bigger Ba atoms are replaced by the smaller Sr atoms, 
which leads to a larger deformation of the surrounding lattice and hence to a larger modulation amplitude. The more homogeneous the filling of these channels with one atomic type (Ba) the lower the modulation amplitude.  Our results also show that the structure can be described with a two-dimensional incommensurate harmonic modulation. No second order modulation has been observed, both by single crystal diffraction measurements and q-scans. The positional modulation of the Nb atoms is much smaller than that of the oxygen atoms, such that the modulation can be seen as a rotational modulation of almost rigid NbO$_{6}$-octahedra. 
} 

\zusammenfassung{
Wir berichten \"uber den Einfluss des Bariumgehalts auf die modulierte Struktur von \\
 Sr$_{0.61}$Ba$_{0.39}$Nb$_{2}$O$_{6}$ (SBN61) 
und  Sr$_{0.34}$Ba$_{0.66}$Nb$_{2}$O$_{6}$ (SBN34) 
bei Umgebungstemperatur. Unsere Resultate basieren auf Neutronendiffraktionsmessungen an Einkristallen und k\"onnen mit einer modulierten Struktur im (3+2) Raum beschrieben werden.
Unsere Resultate k\"onnen wir mit zwei inkommensurablen harmonischen Modulationen beschreiben.
Die Modulationsamplitude der Positionen der Nb Atome is viel kleiner als jene der Sauerstoffatome. Damit k\"onnen wir die Modulation der  NbO$_{6}$-Oktaeder als Rotationen von starren  K\"orpern betrachten und mit einer harmonischen Modulation beschreiben. Weder mit Einkristallmessungen noch mit qscans konnten h\"oherer harmonische Beitr\"age gefunden werden. 
Mit h\"oherem Bariumgehalt wird die Modulation reduziert.
Je homogener diese Kan\"ale mit einem Atomtyp (Ba) gef\"ullt sind, desto kleiner ist die Amplitude der Modulation. Wenn wir den Sr Gehalt erh\"ohen und die 
gr\"osseren Ba Atome durch die kleineren Sr Atome ersetzen werden, erniedrigt sich die Deformation der Gitterumgebung und f\"uhrt damit zu einer
Reduktion der Modulationsamplitude, wie sie in unserem Experiment beobachtet wird.
Dies ist ein Indiz daf\"ur, dass die partielle Besetzung der Pentagonkan\"ale durch Sr und Ba Atome die Ursache der Modulation ist.}


\schlagwort{Modulation Optische Materialien}

\dedication{}

\received{}
\accepted{}
\volume{}
\issue{}
\class{}
\Year{}

\maketitle

\section{Introduction}

The uniaxial ferroelectric relaxor Strontiumbariumniobate (SBN), Sr$_{x}$Ba$_{1-x}$Nb$_{2}$O$_{6}$ (0.26\,$\leq x\leq$\,0.87)\cite{Ulex:2004},  
is a photorefractive material usable in a variety of optical applications, such as optical data storage \cite{Ford:1992} and data processing \cite{Yeh:1987} 
as well as  optical phase conjugation \cite{Wood:1987}. These applications are based on the high photorefractive sensitivity. 
Additionally the material has large pyroelectric, electro-optic, and piezoelectric coefficients  \cite{Ewbanks:1987, Lines:2001, Neurgaonkar:1987}. 
Furthermore, SBN is a model substance for the investigation of the relaxor type  ferroelectric phase transition, where ferroelectric nanoclusters are stabilized 
by  internal random fields above the critical temperature $T_c$ over a wide temperature range, 
such that the ferroelectric polarization does not decay spontaneously at $T_c$ \cite{Bhalla:1987}. 
This relaxor behavior is well explained by the Random-Field-Ising model for the ferroelectric phase transition. 
Assuming an internal random electric field, all critical exponents could be determined according to the scaling relation 
\cite{Dec:2001,Granzow:2004,Kleemann:2006}. They fulfill the Rushbroke relation and belong to the universal class of the three-dimensional  
Random-Field-Ising model. Note however, that the model underlying the ferroelectric relaxor properties of SBN is a controversial issue, as discussed, 
e.g., in Ref.\,\cite{Scott:2006}. From a structural point of view the space group of the average structure changes from $P-4b2$ to $P4bm$ 
during this high-temperature ferroelectric phase transition, while the incommensurate modulation remains \cite{Schneck:1981,Balagurov:1987}. 

\begin{figure}[t]
\includegraphics[width=7cm,angle=0]{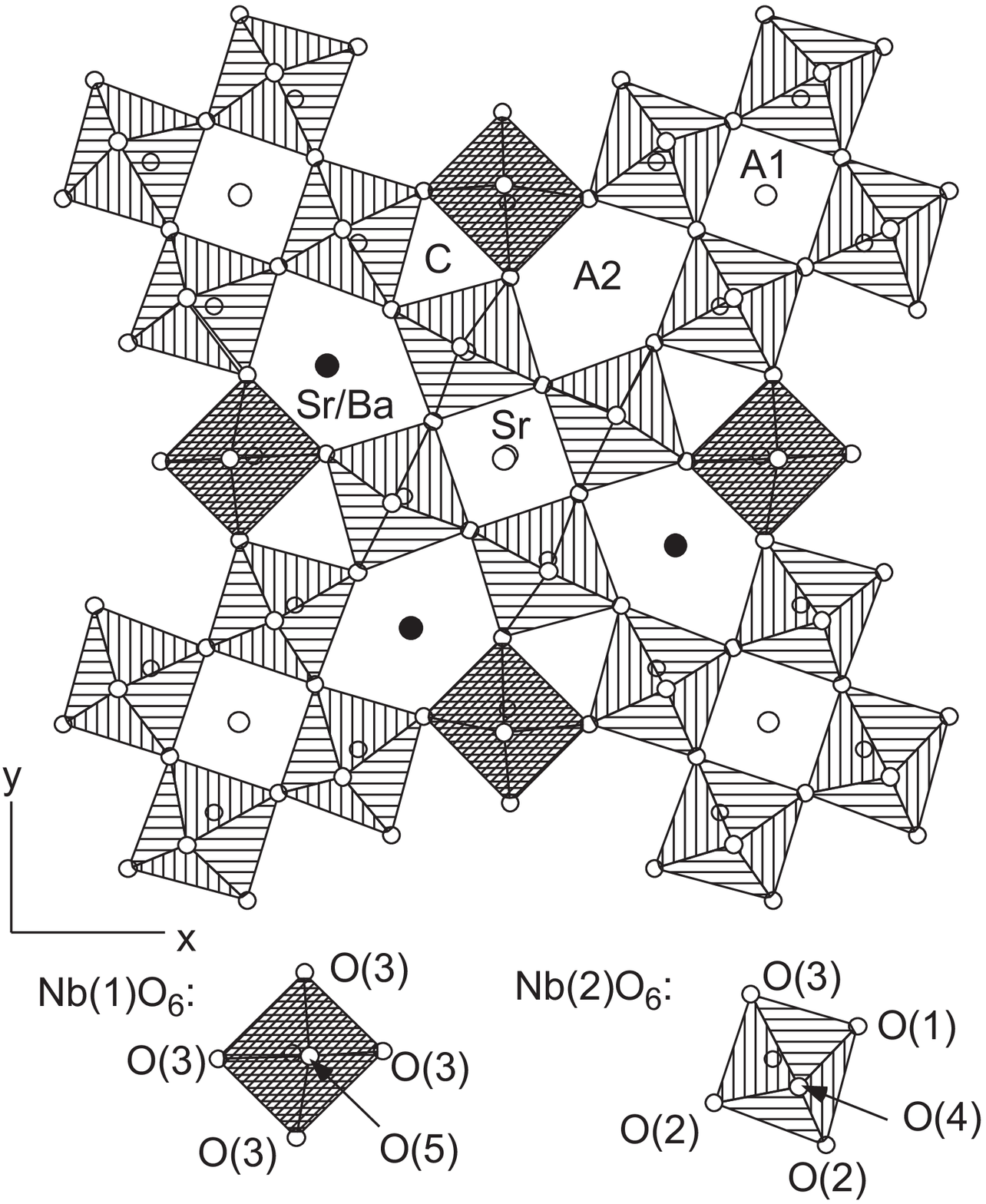}
\caption{Projection of Sr$_{0.61}$Ba$_{0.39}$Nb$_{2}$O$_{6}$ along the c-axis. The pentagonal channels A2 are filled by Strontium and Barium (Sr2/Ba1), the tetragonal 
channels A1 by  Strontium (Sr1) only, and the trigonal channels C remain empty. 5 Sr/Ba atoms are distributed over 6 A1/A2 sites.}
\label{SBN61struktur}
\end{figure}

The average structure of congruently melting Sr$_{0.61}$Ba$_{0.39}$Nb$_{2}$O$_{6}$ (SBN61) in the ferroelectric phase, spacegroup $P4bm$ (No.\,100), 
has been determined by x-ray diffraction \cite{Chernaya:1997}, showing that the structure is a three-dimensional network of NbO$_{6}$-octahedra 
linked at their corners forming alternating five- and four-membered rings (see Fig.\,\ref{SBN61struktur}). The structure contains two 
crystallographically non-equivalent Nb-atoms: Nb1 and Nb2. The Nb(1)-O$_6$-octahedra have point-symmetry $mm2$ and form infinite chains 
of the composition [NbO$_{5}$]$^{5-}$ along the crystallographic $c$-axis. The Nb(2)-O$_6$-octahedra are located in general positions (point symmetry $1$) 
and form channels along the $c$-axis of square cross section. The Sr atoms occupy positions of symmetry $4$ inside these square channels. 
This is illustrated in Fig. \ref{SBN61struktur} showing
the pentagonal channels A1 occupied by Sr only and the channels A2 filled by Sr/Ba. The trigonal channels C remain empty.
The pentagonal channels are wider than the square channels. All the Ba and some of the Sr atoms are located in such larger channels (point symmetry \textit{m}), 
which are not fully occupied. Five Sr and Ba atoms are distributed over six sites. 
In the early structure determinations of SBN \cite{Jamieson:1968} split positions for Oxygen atoms were introduced to account 
for the disorder caused by the Sr/Ba distribution in the crystals.  Later Schneck \textit{et al.}  \cite{Schneck:1981} 
observed satellite spots at positions ($h\,\pm\,\frac{1+\delta}{4}$, $k\,\pm\,\frac{1+\delta}{4}$, $l\,\pm\,\frac{1}{2}$) with $\delta$\,=\,0.26(5) 
in Sr$_{0.71}$Ba$_{0.29}$Nb$_{2}$O$_{6}$, which revealed the incommensurate nature of the structural modulation. These results were confirmed 
by Balagurov \textit{et al.}\,\cite{Balagurov:1987} in neutron time-of-flight measurements on Sr$_{0.7}$Ba$_{0.3}$Nb$_{2}$O$_{6}$ where $\delta$\,=\,0.22(1) 
was found. More time-of-flight studies revealed that the modulation parameter $\delta$ is slightly 
decreasing with decreasing Sr-concentration $x$ for different compositions 0.46\,$<\,x\,<$\,0.75 \cite{Prokert:1991}. 
From electron diffraction measurements on Sr$_{0.5}$Ba$_{0.5}$Nb$_{2}$O$_{6}$ $\delta=0.190(5)$ was reported \cite{Bursill:1987}. 
Only recently x-ray measurements with a systematic collection of satellite reflections were performed on a Sr$_{0.61}$Ba$_{0.39}$Nb$_{2}$O$_{6}$   
single crystal  and subsequently analyzed in terms of the superspace formalism \cite{Woike:2003}.

Recently the average structure of Sr$_{x}$Ba$_{1-x}$Nb$_2$O$_6$  in the composition range 0.32 $<$ x $<$ 0.82 was systematically investigated 
by X-ray diffraction  \cite{Podlozhenov:2006}. It was shown that the lattice parameters a and c decrease with increasing strontium content, 
which could be ascribed to the exchange of Ba by Sr in the pentagonal A2 channels. The occupation of the  square 
A1 channels remains nearly constant for all concentrations x. In order to explore the influence of the composition
 x on the modulated structure and its physical origin in SBN we present here a neutron diffraction investigation on 
single crystals of Sr$_{x}$Ba$_{1-x}$Nb$_{2}$O$_{6}$ with the two compositions x=0.61 and x=0.34. 
As demonstrated in a recent powder diffraction study on SBN \cite{Schefer:2006}, neutron diffraction is especially suited for this 
kind of structural investigation, because the neutron scattering length of O  (5.803 fm) is
 of the same order of magnitude as that of the heavy nuclei Sr (7.02 fm) and Ba (5.06 fm).   
Therefore the description of a positional modulation of the Oxygen atoms beside the heavy atoms 
Strontium and Barium is possible with high accuracy.  A second reason to use neutron diffraction is the fact, that neutrons are 
scattered by the nucleus, making them very sensitive to positional disorder as shown e.g. by many studies investigating Cu-O distances in
high-temperature superconductors as successfully shown by \cite{Hewat:1990}.
The data are analyzed using the superspace approach for the description of the two-dimensional incommensurate modulation.

\section{Experimental and Computational Details}

The crystals of SBN61 and SBN34 were grown by the Czochralski method in the crystal growth laboratory of the University of Osnabr\"uck. \\
For  the q-scans on SBN61 on the cold triple axis spectrometer TASP \cite{Semadeni:2001}/SINQ\cite{fischer:1997} and
the measurements on TriCS\cite{Schefer:2000}/SINQ, a crystal of size 4x4x5mm$^{3}$ was poled by applying an electric field of 500V/mm along the crystallographic c-axis during 6 hours at 23$^{\circ}$C.  The same SBN61 crystal was polarised by applying 270 V/mm at T=130$^{\circ}$C and field-cooled down to room temperature before the full data collection on the hot neutron diffractometer 5C2/LLB (dataset 1). An additional dataset on the same SBN61 crystal has been collected  at TriCS (Tab. VII-IX, appendix \cite{schefer:2008ap})in order to test the crystal for the 5C2/LLB measurement and to make first searches for potential second order satellites.  \\ \\
The SBN61 (dataset 1) crystal has an almost cubic shape with with minor crystallographic faces on the edges. The size of the unpoled crystal of SBN34 measured on TriCS/SINQ (dataset 2) is has maximum dimensions 9x9x9mm$^{3}$  with non
rectangular faces of known crystallographic orientation. All experimental conditions are listed in Table  \ref{tbl:datacollection}. 
\\

\subsection{Data Collection}

For the data collection, we doubled the tetragonal c-axis in order to match the later
refinement with the incommensurate modulation vectors $\mathbf{Q_{1,2}}$=($\alpha$,$\pm \alpha $,$\frac{1}{2}$) when using the original cell. 
This simple transformation with twice the $c$ parameter of the original cell makes it possible to separate internal and external parameters completely. It leads
to an additional centering which can be characterized by one non-primitive centering vector, (0,0,$\frac{1}{2}$,$\frac{1}{2}$,$\frac{1}{2}$).
The  modulated structure satellite reflections are consequently at positions given by the modulation vectors
$\mathbf{Q_{1,2}}$=$\alpha \cdot$($\mathbf{a}^{*} \pm \mathbf{b}^{*})$ with $\alpha$=0.3075  \enspace for SBN61 and $\alpha$=0.2958  \enspace for SBN34, respectively, with  c = 2 $\cdot$c$_{av}$.
Data sets were collected
on the instruments 5C2/LLB (SBN61, dataset 1) and TriCS/SINQ (SBN34, dataset 2). Data collections were performed using $\omega$-scans
and single detectors for all instruments.  Lorenz correction  has been applied to all datasets. 
The measurements are summarized in Table\,\ref{tbl:datacollection}. Absorption was corrected in JANA2000 \cite{Petricek:2000} using the exact shape of the crystal.
\\
The TASP/SINQ instrument was used to perform  high-resolutions q-scans in order to search weak higher order satellites not detected on the single crystal 
diffraction instruments.  TASP is especially suited for this purpose due to its very low background: High collimation and suppressing    inelastic scattering  increase the peak-to-background relation on this instrument dramatically in respect to conventional diffraction instruments such as 5C2 or TriCS.\\

\subsection {Refinement}

The average structure is refined in the space group $P4bm$ (No.\,100). The tetragonal lattice parameters are $c=7.8856(2)$\,\AA \enspace and $a=12.4815(3)$\,\AA \enspace ($c=2 \cdot c_{av}$ has been doubled). The Ba content is fixed at  $1-x=0.39$ for SBN61 and  $1-x=0.66$ for SBN34. 
These values were  determined by X-ray fluorescence \cite{Ulex:2004} and neutron activation \cite{Woike:1997} analysis.
 The sum of the occupancies of atoms Sr1, Sr2 and Ba2 was constraint to 1.  The coordinates as well as the displacement parameters of the Atoms Sr2 and Ba2, sharing the same crystallographic site (4c) in channel A2 in a statistical manner, are constrained. Isotropic extinction correction Type I (Lorentz distribution of mosaics) has been applied \cite{becker:1974a,becker:1974b,becker:1975}. For the modulated structures, the superspace group P4bm 
($\alpha ,\alpha, \frac{1}{2}$,$\alpha - \alpha ,\frac{1}{2}$) as tabulated by De Wolff  \cite{Wolff:1981} has been used. The refinement has been done in 5-dimensional superspace as described by de Wolff  \cite{Wolff:1981} and Janner \& Janssen \cite{Janner:1980a,Janner:1980b} using the program JANA2000 \cite{Petricek:2000}.  The details of this concept for the present case (2 modulation vectors, 5 dimensional space) are given in Refs. \cite{Woike:2003} or \cite{Schaniel:2002}.

\section{Results}
Main structural results are discussed on the example of SBN61, since the peculiarities of the modulated structure are more pronounced than in SBN34. 
 Furthermore the results can be compared to the available X-ray refinement \cite{Woike:2003}. 
 On the other hand, the comparison to the results of the SBN34 structural analysis gives clear evidence of the origin of the modulation.
 In order to clarify a possible disorder of the Strontium and Barium sites and to estimate the influence of anisotropic temperature factors we have also analysed the average structure of SBN61 carefully. This procedure is necessary for a sound interpretation of the subsequent modulated refinements.

\subsection{Average Structure of SBN61}

\begin{figure}[t]
\includegraphics[width=6cm]{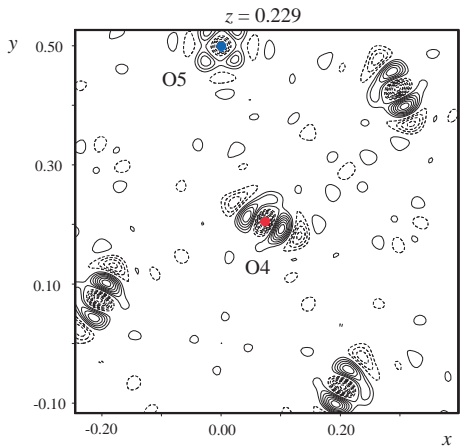}
\caption{Difference Fourier map of the refined average structure of SNB61 at ambient temperature at $z$\,=\,0.229 showing the atoms O4 (center, red color online) (0.0767,0.2049,0.229)   and O5 (top left, blue color online) (0,$\frac{1}{2}$,0.231) (circles). 
Contours 1 , maximum 8.2, minimum -8.2 (fm/ \AA$^3$). 
\label{sbn61-difference-fourier-o4-average} }
\end{figure}

The average structure has been refined using a fixed Ba content of  $1-x=0.39$.
The tetragonal lattice parameters are a=12.481(8)\,\AA, c=7.885(6)\,\AA\enspace (also in this case, we used $c=2 \cdot c_{av}$
for better comparison of the results).
The refinement with isotropic displacement parameters $U_{\mathrm{iso}}$ yields agreement factors $R=0.195$ and Goodness of fit $S=26.3$. The refined parameters are given in  \cite{Schaniel:2003t}. The refinement can be improved by introducing anisotropic displacement parameters $U_{ij}$ ($i,j=1,2,3$) yielding $R=0.093$, $S=12.2$,  
at the cost of negative values for the atom Nb1  (Tables VII and X in the appendix \cite{schefer:2008ap}). In both cases large displacement parameters are obtained for Sr2/Ba2 and all Oxygen atoms. Sr2/Ba2, O4 and O5 exhibit large values of $U_{11}$, $U_{22}$, and $U_{12}$ whereas for O1, O2 and O3 large values of $U_{33}$ are obtained, indicating that the modulation of the former are within the tetragonal plane, whereas that of the latter are along the crystallographic $c$-direction. Refining the atoms Sr2/Ba2  unrestricted does not improve the quality of the fit and is therefore not considered for the final refinement. Looking at the difference Fourier map (Fig.\,\ref{sbn61-difference-fourier-o4-average}) one can clearly see the mismatch in the refinement at the O4 and partially O5 positions for SBN61. Refining split positions for the O4 atom improves the agreement factors significantly to $R=0.063$ but at the cost of negative displacement parameters for several atoms and is therefore discarded as  an unphysical solution of the structure. As is evident from the difference Fourier map (Fig.\,\ref{sbn61-difference-fourier-o4-average}), one would have to introduce more than two positions to describe the behavior at the O5 position.The refinement of the average structure from neutron data is in agreement with that observed by x-rays \cite{Woike:2003}, but also shows that the inclusion of a modulation as done in the next section is imperative.

\subsection{Modulated Structures of SBN61 and SBN34}

The analysis of the modulated structure is done using the superspace approach. Thereby the incommensurate modulation is described using a five-dimensional space, in order to account for the two modulation vectors $\mathbf{Q_{1,2}}$=($\alpha$,$\pm\alpha$,0), where $\mathbf{Q_{1,2}}$ is in respect to the doubled $c$ axis as defined in Tab. \ref {tbl:datacollection}. The method is summarized in detail in Ref. \cite{Woike:2003, schefer:2008ap}.

\begin{figure}[t]
\includegraphics[width=8cm,angle=270]{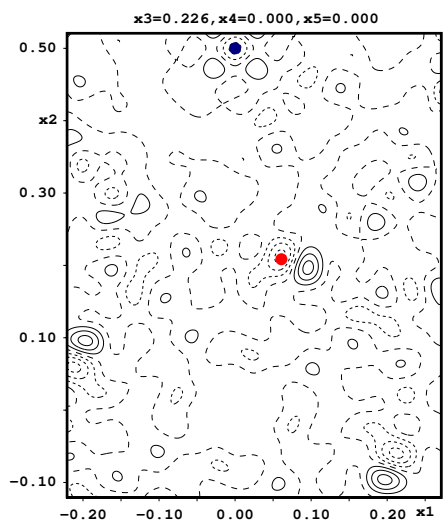}
\caption{Difference Fourier map of the refined modulated structure of SNB61 (dataset 1) at ambient temperature at $z$\,=\,0.225 showing the atoms O4 (center, red online version)  and O5 (top, blue online version). 
Contours maximum 7.3, minimum -7.6, in steps of 2 (fm/ \AA$^3$).
\label{sbn61-difference-fourier-o4-modulated} }
\end{figure}

\begin{figure}[t]
\includegraphics[width=8cm,angle=270]{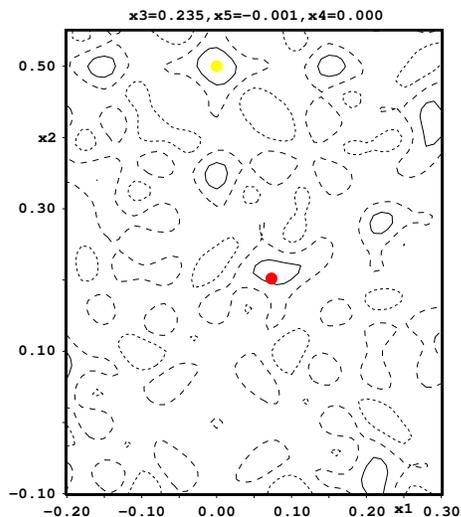}
\caption{Difference Fourier map of the refined modulated structure of SNB34 (dataset 2) at ambient temperature 
at $z$\,=\,0.234 showing the atoms O4 (center, red online version)   and O5 (top, yellow online version). 
Contours 1 , maximum 2.0, minimum -1.9. 
\label{sbn34-difference-fourier-o4-modulated} }
\end{figure}

The refinements are performed in the superspace group 
$P4bm(\alpha \alpha \frac{1}{2},\alpha - \alpha \frac{1}{2})$. 
Starting from the average structure, two harmonic positional modulation waves are introduced for all atoms. Again, the parameters of Sr2 and Ba2 located in channel A2 (see Fig. \ref{SBN61struktur}) are constrained to have identical positional and displacement parameters in all refinements. 
Introduction of an additional occupational disorder function for  Sr2/Ba2 did not result in significant improvements of the agreement factors and was therefore not taken into account in the final refinement. However we introduced two modulation waves for the displacement parameters of the Sr2/Ba2 atoms in order to take into account that two different atoms occupy the same position. These two modulation waves then incorporate effects originating from the slightly varying positions of Sr2/Ba2 in the disordered lattice, e.g., different orientations of the thermal ellipsoids.  The displacement parameters are chosen anisotropic, yielding minor negative values for the Oxygen atoms. 
 The refinement for the two datasets of  SBN61 collected on the instrument  5C2 (dataset 1) and TriCS (dataset 3 with lower q-range, discussed in detail in the thesis of Schaniel \cite{Schaniel:2003t}, tables summarized in the appendix \cite{schefer:2008ap}) are in excellent agreement. However, small residual differences remain, as e.g shown by the difference Fourier maps  around oxygen atoms O4 and O5 (Fig.\,\ref{sbn61-difference-fourier-o4-modulated}), but are dramatically reduced compared to the refinement using an average structure (see Fig.\,\ref{sbn61-difference-fourier-o4-average}).  As illustrated in Fig.\,\ref{sbn34-difference-fourier-o4-modulated} similar difference Fourier maps are obtained for SBN34 after final refinement, however the remaining residuals are much smaller than in the SBN61.

 The final refinement of the structure yields total agreement factors of $R_{w,obs}=0.0908$, $S=7.15$ and agreement factors of $R_{\mathrm{all}}=0.0579$ for main  and $R_{\mathrm{w,all}}=0.192$ for satellite reflections for SBN61.  For SBN34 total agreement factors of $R_{w,obs}=0.111$, $S=6.96$ and agreement factors of $R_{\mathrm{all}}=0.091$ for main  and $R_{\mathrm{w,all}}=0.134$ for satellite reflections are obtained. This is approximately a factor of two above the agreement factors obtained from the extended X-ray data sets of SBN61 collected by Woike et al. \cite{Woike:2003} yielding a R$_w$ of 0.045 for the main reflections and R$_w$ of 0.126 for the satellite reflections. This indicates that the positions of the oxygen could have a systematic error, as  neutrons are more sensitive to the oxygen positions compared to X-ray diffraction studies. Note however that the X-ray study on SBN61 was performed only up to $\sin{\theta}/\lambda$-values of 0.75 \AA$^{-1}$, while our neutron study goes up to 1.0 \AA$^{-1}$.

\begin{figure}[t]
\includegraphics[width=8cm,angle=270]{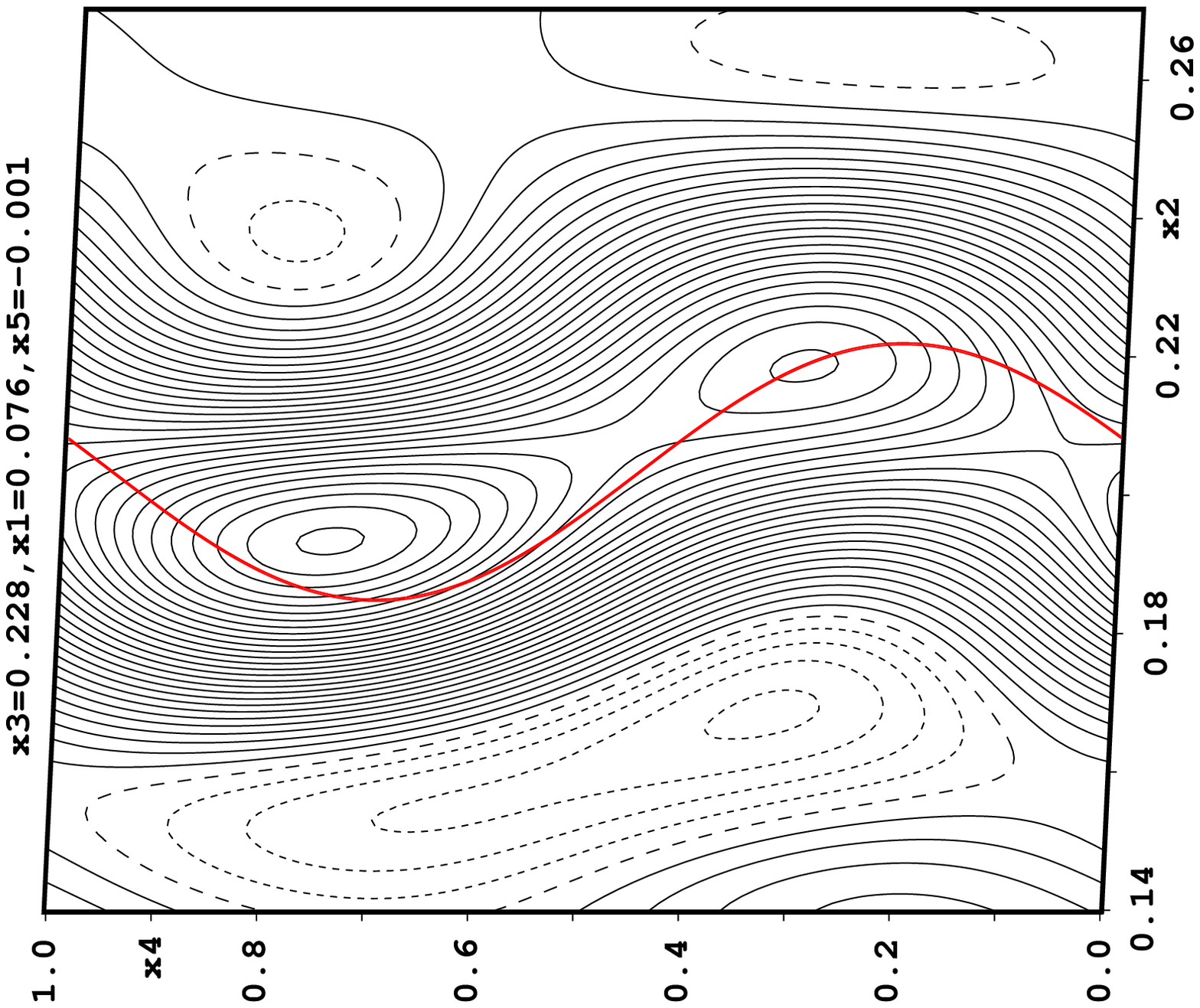}
\caption{Observed positional modulation of Oxygen atom O4 of SBN61 at ambient temperature. 
Lines (color online) are fitted atomic positions. The contours are from -8.5 to 56.4 in  steps of 2 (fm/ \AA$^3$). 
\label{modulationobs-sbn61-o4}}
\end{figure}

\begin{figure*}[t]
\includegraphics[width=8cm,angle=270]{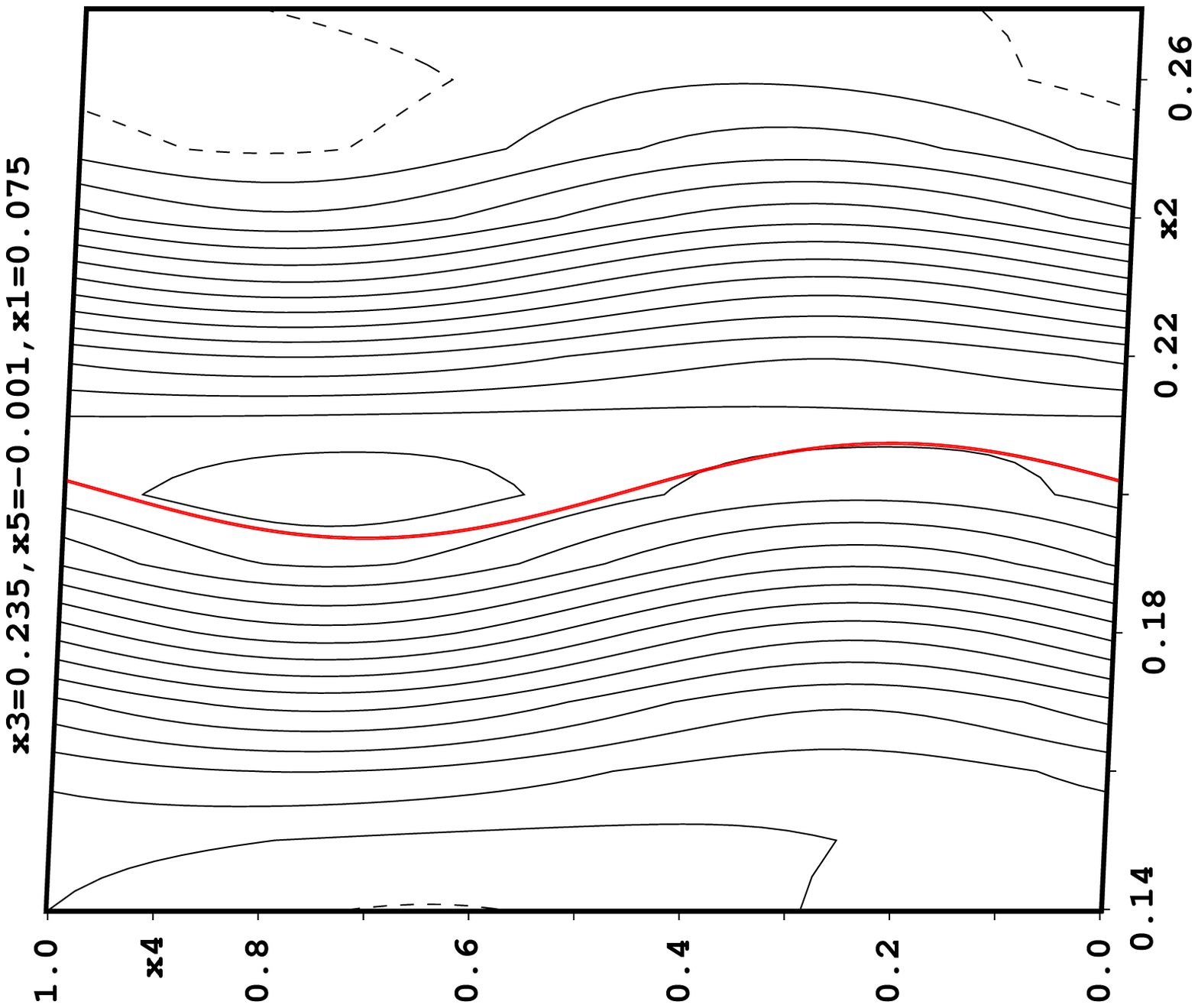}
\caption{Observed positional modulation of Oxygen atom O4 of SBN34 at ambient temperature.
Lines (color online) are fitted atomic positions. The contours are from -1.4 to 31.4 in  steps of 2 (fm/ \AA$^3$).} 
\label{modulationobs-sbn34-o4}
\end{figure*}

 The full parameter set of the final refinement are available from the supplementary material \cite{schefer:2008ap}. It can be seen that the form of the modulation is the same for SBN61 and SBN34, but the amplitude is reduced by a factor of two in SBN34. A comparison between selected values is listed in  Table \ref{tbl:interatomic-distances-comparison}. The sine- and cosine part of the harmonic modulation waves belonging to the two modulation vectors are indicated by $s,m,n$ and $c,m,n (m,n=0,1)$, \textit{e.g.} $s,1,0$ is the sinusoidal modulation contribution from the modulation vector $\mathbf{Q_{1}}$.  Overall the atoms Sr2/Ba2, O4, and O5  are modulated in the tetragonal $ab$-plane whereas the atoms O1, O2, and O3  are mainly modulated along the $c$-direction of the crystal. The atoms Nb1, Nb2, and Sr1  exhibit only a small positional modulation. Taking exemplarily the atom O4 (which has the highest modulation amplitude) for comparison of SBN61 and SBN34, we find that the s,1,0 amplitudes are reduced by a factor of 3 and the c,1,0 amplitudes by a factor of 2. This difference in modulation amplitude is then also reflected in the minimal/maximal distances (selected values listed in Tab. \ref{tbl:interatomic-distances-comparison}, full listing 
 see Tables III/IV and VII/VII of the supplementary material \cite{schefer:2008ap}), where for Nb2-O4 min/max distances of 1.721(14)\,\AA/2.029(15)\,\AA \enspace and 1.746(19)\,\AA/2.010(19)\,\AA \enspace are found for SBN61 and SBN34, respectively. 
For Nb2-O4$^{iv}$ the difference is even more pronounced with distances of 1.959(14)\,\AA/2.213(15)\,\AA \enspace and 2.022(19)\,\AA/2.280(19)\,\AA \enspace for SBN61 and SBN34, respectively. In both cases the average distances, 1.889(15)\,\AA/1.87(2)\,\AA \enspace for Nb2-O4 and 2.109(15)\,\AA/2.13(2)\,\AA \enspace for Nb2-O4$^{iv}$, are the same within experimental error for SBN61 and SBN34. 
Compared to the results of the X-ray study for  SBN61\cite{Woike:2003}, where the interatomic distances Nb2-O5 and Nb2-O5$^{iv}$ (O4 and O5 are interchanged in the two papers) range from   1.79(3)\,\AA \enspace to 1.98(3)\,\AA \enspace, our study shows values between 1.95(3)\,\AA \enspace and  2.18(3)\,\AA.
Hence the modulation of the Nb2-O4/O4$^{iv}$ distances in SBN61 of 0.30(3)\,\AA/0.25(3)\,\AA \enspace in the neutron case is significantly larger than the 0.19(6)\,\AA/0.23(6)\,\AA \enspace found in the X-ray study. This fact is also nicely observed in the Fourier maps, as shown for the observed positional modulation of O4,  together with the fitted position along the first modulation vector $\mathbf{Q_{1}}$ in SBN61 in Fig.\,\ref{modulationobs-sbn61-o4}, which shows the larger modulation amplitude compared to the corresponding figure in the X-ray study (Fig.\,6 in Ref.\,\cite{Woike:2003}). From these figures it becomes also clear why the agreement factors in the neutron case are worse than in the X-ray case, although the description of the O4 modulation seems slightly better: the high oxygen sensitivity of the neutron yields high R-values even for small mismatches in the modulation description. In SBN34, as illustrated in Fig.\,\ref{modulationobs-sbn34-o4}, the modulation amplitude is significantly reduced as discussed above, leading to better agreement factors. 
 
\begin{figure}[t]
\includegraphics[width=10cm,angle=0]{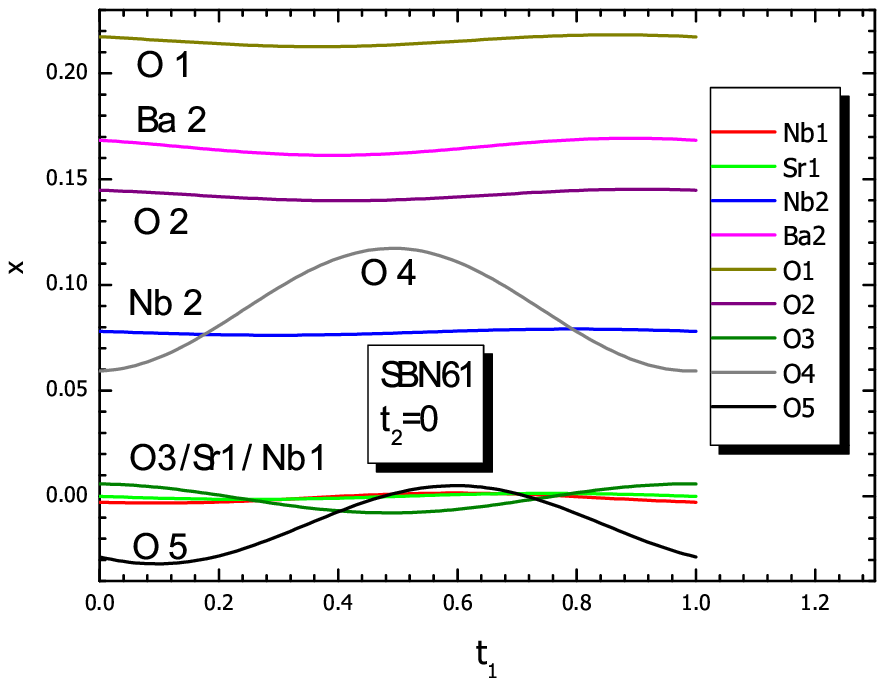}
\caption{Modulations of x component from  the atoms Nb1, Nb2, Sr1, Ba2, and 01-O5 in SBN61  along $t_1$ at $t_2$=0.}
\label{sbn61-modulation-t}
\end{figure}

\begin{figure}[t]
\includegraphics[width=10cm,angle=0]{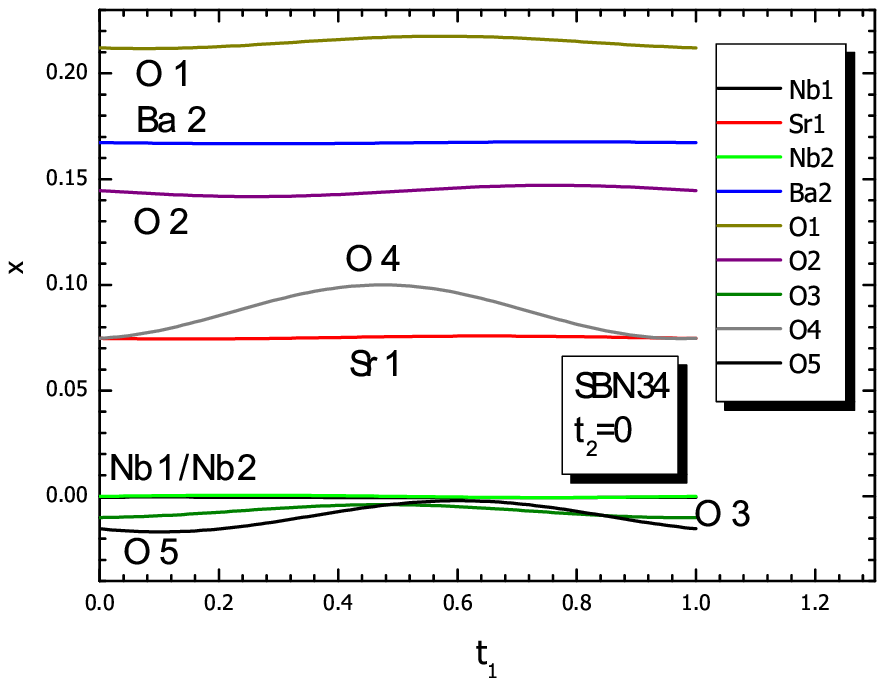}
\caption{Modulations of x component from the atoms Nb1, Nb2, Sr1, Ba2, and 01-O5 in SBN34  along $t_1$ at $t_2$=0.}
\label{sbn34-modulation-t}
\end{figure}

\begin{figure}[t]
\includegraphics[width=8cm,angle=270]{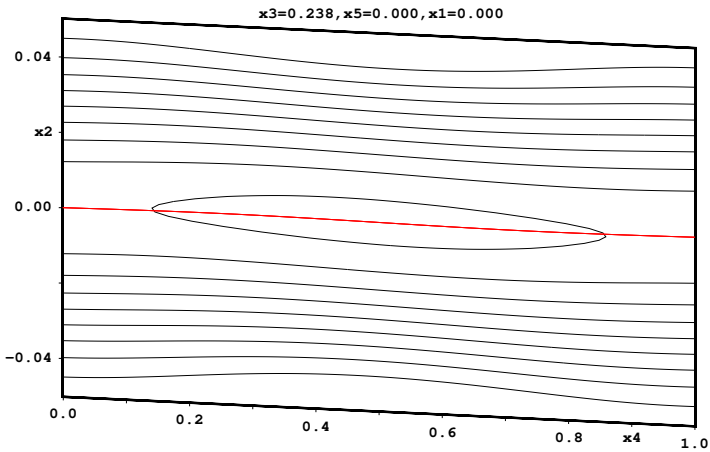}
\caption{Observed  Fourier  map in the x2/x4 plane at ambient temperature through the center of the Sr1 atom, 
showing the absence of any modulation for the Strontium atom Sr1.
The bold horizontal line (red color online) is the fitted atomic position of Sr1. 
The contours are from -1.36 to 18.53 in  steps of 2 (fm/ \AA$^3$).
\label{modulationobs-sbn61-sr1}}
\end{figure}

Similar observations concerning average, minimal, and maximal values of distances are made for the Sr/Ba-polyhedra, where in the neutron refinement a larger amplitude is found than in the X-ray refinement, especially for the distances including the atoms O4 and O5. Here again the Sr/Ba atoms have a much smaller modulation amplitude than the oxygen atoms, as illustrated exemplarily for SBN61 in Fig.\,\ref{sbn61-modulation-t}, which shows how the x-coordinate is changing as a function of t$_{1}$ for atoms Ba2, Sr1, Nb1, Nb2, ad O4,
where x is defined as 

\begin{eqnarray*}
x = x_{o} + \\
U_{xs,1,0} \cdot sin(2  \pi \cdot \mathbf{Q_{1}} \cdot \mathbf{r_{0}}+t_{1}) + \\
U_{xc,1,0} \cdot cos(2  \pi \cdot \mathbf{Q_{1}} \cdot \mathbf{r_{0}}+t_{1}) + \\
U_{xs,0,1} \cdot sin(2  \pi \cdot \mathbf{Q_{2}} \cdot \mathbf{r_{0}}+t_{2}) + \\
U_{xc,0,1} \cdot cos(2  \pi \cdot \mathbf{Q_{2}} \cdot \mathbf{r_{0}}+t_{2})  
\end{eqnarray*}
, where U$_{yz,m,n}$ are  the modulation amplitudes listed in Tab. \ref{tbl:modulation-SBN61-LLB-SBN34-TriCS}.\\
This means, for $t_{2}=0$ the third and fourth term make just a constant but generally non-zero contribution to the  curves shown in Fig. \ref{sbn61-modulation-t} for SBN61 and
Fig. \ref{sbn34-modulation-t} for SBN 34, respectively.
The Sr1 atoms in the tetragonal A1 channels are hardly modulated as shown in 
Fig. \ref{modulationobs-sbn61-sr1}. Corresponding modulations in SBN34 are drastically reduced (Fig. \ref{sbn34-modulation-t}).
Detailed  interatomic distances and full tables are listed in the supplementary material \cite{schefer:2008ap}. 
\\
It is known from x-ray diffraction experiments on the average structure, that the occupation of the square channels (Sr1) 
is hardly affected by the decreasing Sr content \cite{Podlozhenov:2006}. The exchange of Sr and Ba takes place in the pentagonal channels. 
So for the SBN61 the site occupancies (in comparison to the the maximum occupancy of 1) are 0.72 for Sr on A1 (2a), 0.4875 for Ba on A2 (4c), and 0.402 for Sr on A2 while for SBN34 the corresponding site 
occupancies are 0.62 for Sr on A1, 0.825 for Ba on A2, and 0.11 for Sr on A2 \cite{Podlozhenov:2006}. Therefore in SBN34 we have an almost uni-atomic 
occupancy of the two sites, the A1 channels are solely occupied by Sr while on the A2 sites the Sr:Ba ratio is 1:7 compared to 6:5 in SBN61. 
Hence we can argue that the reduced amplitude of the modulation is due to the reduced distortion exerted on the lattice by the 
more homogeneous filling ratio by only one atomic type (Ba) of the A2 channels in SBN34. The modulation does not disappear due to 
the fact that still one of six sites remains empty. 

\begin{figure}[t]
\includegraphics[width=12cm,angle=0]{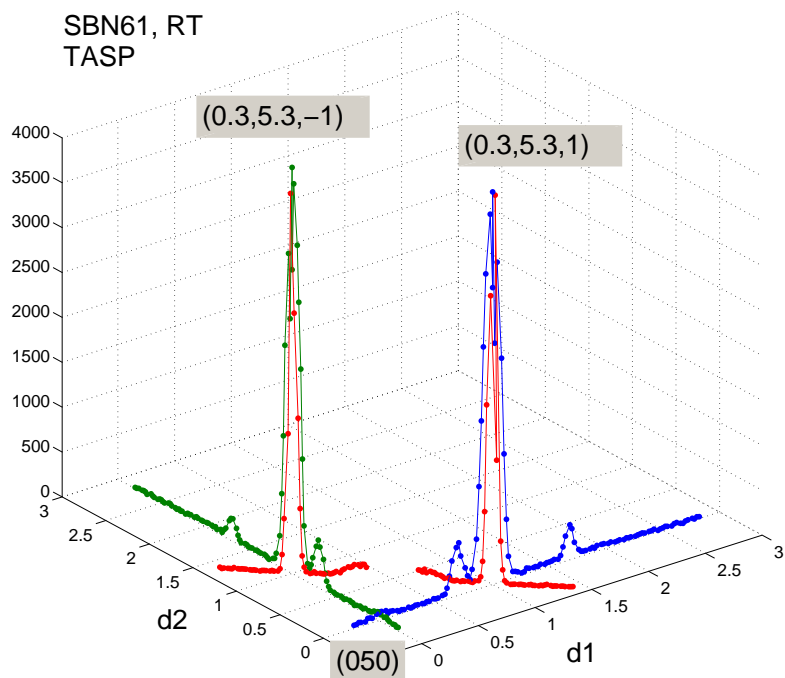}
\caption{Q-scans of SBN61  in a plane formed by the (050) reflection  and the two reciprocal 
vectors  $\mathbf{d_{1}}$=(.3075,.3075,1) and $\mathbf{d_{2}}$=(.3075,.3075,-1) with respect to the double c-axis as
given in table \ref{tbl:datacollection}. Measurement performed
on TASP/SINQ with ($\lambda$=2.3603\AA) at ambient temperature.
Side reflections of the satellites correspond to 
(002)- and (111)-aluminum-powder lines 
from holder of the crystal, crossed by the 2  scans along  $\mathbf{d_{1}}$ (color online: blue) and  $\mathbf{d_{2}}$ 
(color online: green) and the (111)-Al line touched by the scan along  $\mathbf{d_{2}}$-$\mathbf{d_{1}}$(color online: red)
\label{sbn61-tasp-contour}}
\end{figure}

Both, triple-axis experiments (q-scans on TASP/SINQ, Fig.  \ref{sbn61-tasp-contour}) and measurements on the single crystal instruments 5C2/LLB (test of 829 2$^{nd}$-order satellites) and TriCS (150 2$^{nd}$-order satellites) showed no evidence for higher order satellites in SBN61.
Therefore only harmonic modulation waves have been introduced for our refinement. There is obviously no satellite observed for positions d$_{1,2}=2$ in Fig. \ref{sbn61-tasp-contour}. The observed side reflections  of the satellites (0.3075,5.3075,-1) and (0.3075,5.3075,1) correspond to 
(002)- and (111)-aluminum-powder lines from the sample holder.

\section{Conclusions}
The structures of Sr$_{x}$Ba$_{1-x}$Nb$_2$O$_{6}$, x=0.61 and 0.34, 
at ambient temperature can be described  with two incommensurate   
modulations vectors $\mathbf{Q_{1,2}}$=($\alpha,\pm \alpha$,0) ($\alpha$=0.3075 for SBN61 and $\alpha$=0.2958 for SBN34) in a harmonic approximation. 
Our study clearly shows, that the modulation amplitude of the oxygen O4 atoms is decreasing by a factor of 2 going from SBN61 to SBN34 (increased Ba content), but has the modulation has the same shape. The positional modulation of the Nb atoms is much smaller than that of the oxygen atoms. It is  is therefore originating mainly from a rotational 
modulation of the NbO$_{6}$-octahedra. The physical origin for this rotational modulation is most probably the filling of the pentagonal 
A2 channels by Sr and Ba atoms, as evidenced by the decrease of the modulation amplitude with increasing Ba content when going from SBN61 to SBN34. Due to their different size the surrounding lattice is deformed leading to the modulation of the oxygen atoms.  The atom with the highest modulation is the apical O4 atom, which is in the same plane as the Sr/Ba atoms. The second type of NbO$_6$ octahedra is significantly 
less influenced by the modulation, as seen by the lower modulation amplitude of O5 (Nb(2)-O$_6$-octahedra) compared to O4 (Nb(1)-O$_6$-octahedra). 
 Remaining intensities in the difference fourier maps around O4/O5 as well as slightly negative temperature factors show, that the model needs further improvement. However, without observing higher order satellites refinement has to stay in the harmonic approximation. From  our diffraction data and extended q-scans, we can exclude  satellites of second order, such as $2\cdot \mathbf{Q_1}$, $2\cdot \mathbf{Q_2}$,
 $\mathbf{Q_1}+\mathbf{Q_2}$,  $\mathbf{Q_1}-\mathbf{Q_2}$. Also  extended q-scans did not show any higher order satellites.

\section {Acknowledgment}
Neutron beam  time on the instruments TriCS and TASP  of the Laboratory for Neutron Scattering (LNS) 
at the Swiss Spallation Neutron  Source SINQ, Villigen, Switzerland, and the instrument 5C2 of the Laboratoire L\'eon Brillouin (LLB) at 
the Orph\'ee reactor in Saclay, France, is gratefully acknowledged as well as the support of the crystal growth department of the University of 
Osnabr\"uck. The development of the JANA2000 program package was supported by the Grant Agency of the Czech Republic, grant 202/06/0757.


\clearpage

\begin{table*}
\caption{Experimental data collection parameters for Sr$_{x}$Ba$_{1-x}$Nb$_{2}$O$_{6}$, SBN61 (x=.61) and SBN34 (x=.34), at ambient temperature. The detailed results are listed in a separated appendix \cite{schefer:2008ap}.} 
\label{tbl:datacollection} {\smallskip}
\begin{small}
\begin{tabular}{lrr}
 	 & 5C2/LLB  & TriCS/SINQ\\           \hline
 	 &  SBN61 & SBN34 \\
 	 Dataset & 1 & 2 \\
 	 space group & $P4bm(\alpha,\alpha,\frac{1}{2},\alpha-\alpha,\frac{1}{2}$) 
 	                                                                            & $P4bm(\alpha,\alpha,\frac{1}{2},\alpha-\alpha,\frac{1}{2}$)
 	 \rule{0ex}{3ex} \\
	Z                             & 10  & 10\\
	Radiation                     & n & n \\
	wavelength $\lambda$\,(\AA)   &  0.835(1) & 1.1800(13) \\
	$T$\,(K)                      &  300 & 300   \\
	 a=b (\AA)                    & 12.4815(3) & 12.4968 (30)   \\
	 c = 2 $\cdot$c$_{av}$ (\AA)  &  7.8856(2) & 7.9604(20)  \\

$\mathbf{Q_{1,2}}$=$\alpha \cdot$($\mathbf{a}^{*} \pm \mathbf{b}^{*})$ 	 , $\alpha$=          
	 	                            & 0.3075  & 0.2958 \\
	 V (\AA$^{3}$)                & 1228.5 & 1243.2 \\
	 d (g/cm$^{3}$)               & 5.256  & 5.371\\
	$[\sin{(\theta)}/\lambda]_{\mathrm{max}}$\,(\AA$^{-1}$)          & 1.00 & 0.694 \\
	 abs. coeff. (mm$^{-1}$)                                         &  0.001  & 0.0018 \\
	 $T_{\mathrm{min}}^{**}$                                         & 0.9944   & 0.9866\\
	 $T_{\mathrm{max}}^{**}$                                         & 0.9953    & 0.9881  \\
	 \hline
	 crystal dimensions $a \cdot b \cdot c$ (mm$^{3}$)                & $4 \cdot 4 \cdot 5$  & $8.7 \cdot 8.5 \cdot 8.92$ \\
	 Polarisation [V/mm], Temperature [$\deg$C]                       &  270, 130 & not poled \\
	 	 crystal  volume (mm$^{3}$)                                     & 80 & 457  \\
	 	 \hline
	 $h_{\mathrm{max}}$                                               &   24  & 14  \\
	 $k_{\mathrm{max}}$                                               & 17    & 15 \\
	 $l_{\mathrm{max}}$                                               & 15    & 9  \\
	 $m_{\mathrm{max}}$                                               & 1     & 1 \\
	 $n_{\mathrm{max}}$                                               & 1     & 1 \\
	
	 no. of refined reflections                                        & 5256   & 1236   \\
	 no. of obs.\,reflections\,(I$>$3$\sigma$)                         & 2527   & 3620   \\
	 no. of obs.\,main reflections\,(I$>$3$\sigma$)                    & 1220   & 459   \\
	 no. of obs.\,first order satellite reflections\,(I$>$3$\sigma$)   & 1307   & 777   \\
	 no. of obs.\,second order satellite reflections\,(I$>$3$\sigma$)  & 0      & -    \\ 
	 	 no. of measured\,second order satellite reflections\            & 829    & -     \\ 
	 $R_{\mathrm{int}}^{\dagger}$                                      & 0.0285 & 0.015 \\
	 $g_{\mathrm{iso}}^{\ddagger}$\,(10$^{-4}$)                        & 0.092  & .005  \\
	 \hline
	 {Refinement}$^1$ \\
	 S                 &  7.39           &   6.96\\
	 R$_{obs}$,R$_{all}$          & 11.56            &  9.06\\
	 R$_{w,obs}$,R$_{w,all}$         &  9.08            &  11.16  \\

	 \hline
	  	main reflections   & &  \\

	R$_{obs}$, R$_{all}$          &   5.79           &  7.15                        \\
	R$_{w,obs}$, R$_{w,all}$          &   6.38           &  10.39                    \\
	\hline
	satellites of order 1   & & \\

  R$_{obs}$,R$_{all}$            &      21.24      &   13.86                     \\
	R$_{w,obs}$, R$_{w,all}$         &      19.23      &   13.46                  \\
	\hline
	  
           \hline
  \multicolumn{3}{l}{       $^{**}$transmission factors, \textit{i.e.} minimal and maximal amount of transmitted neutrons}\\
	\multicolumn{3}{l}{$^{\dagger}$$R$-\textit{factors} of merging process; $^{\ddagger}$Isotropic extinction correction of type I (Lorentzian distriution) is used \cite{becker:1974a,becker:1974b,becker:1975}.}\\
 \multicolumn{3}{l}{${^1}$: refinement using isotropic temperature factors, all agreement factors in [\%].}\\
\end{tabular}
\end{small} 
\end{table*}

\begin{table*} 
\caption{Atomic Fourier amplitudes of selected atoms in SBN61/LLB (dataset 1)
and SBN34/TriCS (dataset 3) at ambient temperature of the displacive modulation functions (details of the formalism are given in Eq. 3.24 in  \cite{Schaniel:2002}. The full table is given in an appendix \cite{schefer:2008ap}. Wave symbol c correspond to a cosinus, s to a sinus modulation, 1 and 2 to modulation vectors $\mathbf{Q_{1,2}}$) 
 } 
\label{tbl:modulation-SBN61-LLB-SBN34-TriCS}
\begin{small} 
\begin{tabular}{lc|ccc|ccc}
\hline  
atom &     wave        &   x         & y           &  z        & x & y & z \\   
\hline
 & & \multicolumn{3}{c}{SBN61, 5C2, dataset 1} & \multicolumn{3}{c}{SBN34, TriCS, dataset 2}  \\
Nb1 &       &  0           &  0.5         &  0.0019(5)    &   0          &  0.5         &  0.0067 \\
    & s,1,0 & -0.0024(4)   & -0.0024(4)   &  0            &              &  0.0001(6)   &  0.0001(6)     \\
    & c,1,0 &  0           &  0           & -0.0034(10)   &   0          &  0           &  0.0004(10)    \\
    & s,0,1 &  0.0010(5)   & -0.0010(5)   &  0            & 0.0006(6)    & -0.0006(6)   &  0            \\
    & c,0,1 &  0           &  0           & -0.0017(10)   & 0            &  0           & -0.0022(10) \\
Nb2 &       &  0.07463(6)  &  0.21139(6)  & -0.0085(4)    & 0.07397(10)  &  0.21084(10) &  0.0016(6)\\
    & s,1,0 & -0.0010(4)   & -0.0013(4)   & -0.0034(5)    & -0.007(5)    & 0.0016(5)    & -0.0002(5)\\
    & c,1,0 &  0.0011(4)   &  0.0016(4)   & -0.0024(5)    & -0.0007(5)   &  0.0016(5)   & -0.0002(5) \\
    & s,0,1 & -0.0039(4)   &  0.0024(4)   & -0.0019(5)    & -0.0001(5)   &  0.0010(4)   &  0.0002(5)  \\
    & c,0,1 &  0.0021(4)   & -0.0021(4)   &  0.0004(5)    & -0.0003(4)   &  0.0019(4)   & -0.0003(5) \\
Sr1 &       &  0           &  0           &  0.2368(5)    &  0           &  0           &  0.2426(10) \\
    & s,1,0 & -0.0014(6)   &  0.0003(6)   &  0            &  0.0005(6)   &  0.0026(6)   & 0           \\
    & c,1,0 &  0           &  0           & -0.0026(10)   &  0           &  0           & -0.0134(14) \\
    & s,0,1 &  0.0003(6)   &  0.0014(6)   &  0            &  0.0026(6)   & -0.0005(6)   & 0\\
    & c,0,1 &  0           &  0           & -0.0026(10)   & 0            &  0           & -0.0134(14)  \\
Ba1 &       &  0.17215(11) &  0.67215(11) &  0.241        &  0.1729      &  0.6729      &  0.2461 \\
    & s,1,0 &  0.0032(5)   &  0.0032(5)   & -0.0030(14)   &  0.0001(4)   &  0.0001(4)   &  0.0019(14)\\
    & c,1,0 &  0.0024(6)   &  0.0024(6)   &  0.0081(12)   &  0.0004(4)   &  0.0004(4)   &  0.0027(15)\\
    & s,0,1 &  0.0057(10)  & -0.0057(10)  &  0            &  0.0060(5)   & -0.0060(5)   &  0 \\
    & c,0,1 & -0.0038(6)   & -0.0038(6)   &  0.0011(15)   & -0.0016(4)   & -0.0016(4)   &  0.0003(14)\\
Sr2 &       &  0.1721      &  0.6721      &  0.241        &  0.17295(16) &  0.67295(16) &  0.2461(11) \\
    & s,1,0 &  0.0032(5)   &  0.0032(5)   & -0.0030(14)   &  0.0001(4)   &  0.0001(4)   &  0.0019(14) \\
    & c,1,0 &  0.0024(6)   &  0.0024(6)   &  0.0081(12)   &  0.0004(4)   &  0.0004(4)   &  0.0027(15) \\
    & s,0,1 &  0.0057(10)  & -0.0057(10)  &  0            &  0.0060(5)   & -0.0060(5)   &  0 \\
    & c,0,1 & -0.0038(6)   & -0.0038(6)   &  0.0011(15)   & -0.0016(4)   & -0.0016(4)   &  0.0003(14) \\
O1 &        &  0.21835(10) &  0.28165(10) & -0.0214(6)    &  0.21663(14) &  0.28337(14) & -0.0143(9)\\
   & s,1,0  &  0.0003(7)   &  0.0003(7)   &  0            & -0.0029(6)   & -0.0029(6)   &  0\\
   & c,1,0  &  0.0028(8)   & -0.0028(8)   &  0.0186(9)    & -0.0005(7)   &  0.0005(7)   &  0.0111(8)\\
   & s,0,1  & -0.0024(7)   &  0.0024(7)   &  0.0099(10)   & -0.0009(6)   &  0.0009(6)   &  0.0091(8)\\
   & c,0,1  & -0.0032(7)   &  0.0032(7)   & -0.0151(10)   & -0.0021(7)   &  0.0021(7)   & -0.0074(9)\\
O2 &        &  0.13938(10) &  0.06819(9)  & -0.0268(6)    &  0.14047(15) &  0.06960(13) & -0.0209(8) \\
   & s,1,0  & -0.0005(8)   & -0.0023(7)   &  0.0220(8)    & -0.0024(7)   &  0.0010(7)   &  0.0105(6) \\
   & c,1,0  &  0.0027(8)   & -0.0053(7)   &  0.0170(10)   &  0.0011(7)   &  0.0016(7)   &  0.0085(7)\\
   & s,0,1  & -0.0055(8)   & -0.0034(8)   &  0.0157(9)    & 0.0034(7)    & -0.0006(6)   &  0.0081(6) \\
   & c,0,1  &  0.0039(8)   & -0.0036(7)   & -0.0237(8)    &  0.0035(7)   &  0.0015(7)   & -0.0163(6) \\
O3 &        & -0.00585(10) &  0.34357(9)  & -0.0280(6)    & -0.00618(15) &  0.34372(15) & -0.0186(9) \\
   & s,1,0  &  0.0037(7)   &  0.0041(6)   & -0.0291(6)    & -0.0014(7)   & -0.0055(6)   & -0.0127(6) \\
   & c,1,0  &  0.0058(8)   &  0.0017(8)   & -0.0093(10)   & -0.0027(7)   &  0.0012(7)   & -0.0041(7) \\
   & s,0,1  & -0.0041(6)   & -0.0052(5)   & -0.0256(8)    &  0.0026(7)   &  0.0009(6)   & -0.0184(7) \\
   & c,0,1  &  0.0030(9)   & -0.0008(8)   &  0.0094(11)   &  0.0009(7)   & -0.0013(7)   &  0.0050(6) \\
O4 &        &  0.07605(18) &  0.20492(15) &  0.2275(5)    &  0.0749(2)   &  0.20557(19) &  0.2351(9)\\
   & s,1,0  & -0.0137(6)   &  0.0178(3)   &  0.0041(6)    & -0.0044(5)   &  0.0067(4)   & -0.0086(10)\\
   & c,1,0  & -0.0256(4)   &  0.0050(4)   & -0.0062(5)    & -0.0119(5)   &  0.0014(4)   & -0.0032(10)\\
   & s,0,1  & -0.0087(7)   &  0.0016(6)   &  0.0091(8)    & -0.0107(5)   &  0.0056(4)   & -0.0069(9)\\
   & c,0,1  &  0.0104(6)   & -0.0014(6)   & -0.0024(9)    &  0.0101(5)   & -0.0049(4)   &  0.0017(11)\\
O5 &        &  0           &  0.5         &  0.2301(6)    &  0           &  0.5         &  0.2347(11)\\
   & s,1,0  & -0.0185(6)   & -0.0185(6)   &  0            & -0.0074(6)   & -0.0074(6)   &  0 \\
   & c,1,0  &  0           &  0           &  0.002(2)     &  0           &  0           & -0.002(2) \\
   & s,0,1  &  0.0163(4)   & -0.0163(4)   &  0            &  0.0117(6)   & -0.0117(6)   &  0 \\
   & c,0,1  &  0           &  0           & -0.0124(16)   &  0           &  0           &   -0.001(2) \\
\noalign{\smallskip}
\hline 
 	 \end{tabular}
 	 \end{small} 
 	 \end{table*}

\clearpage

\begin{table}
\caption{Symmetry codes used in JANA2000.} 
\label{tbl:pbnm-mod-symmetry}
\begin{tabular}{cccccc}  
\hline
(i)    & $ x1$               &  $x2$                 &   $x3$     &   $x5$     &    $-x4$  \\
(iii)  & $\frac{1}{2}-x1$    &  $\frac{1}{2}+x2$     &   $x3$     &  $-x5$     &    $-x4$  \\
(iv)   & $-x1$               &  $-x2$                &   $x3$     &  $-x4$     &    $-x5$  \\
(v)    & $\frac{1}{2}-x2$    &  $\frac{1}{2}$-x1     &   $x3$     &  $-x4$     &    $ x5$  \\
(vi)   & $ x2 $              &  $-x1$                &   $x3$     &  $-x5$     &    $ x4$  \\
(vii)  & $\frac{1}{2}+x$1    &  $\frac{1}{2}-x2$     &   $x3$     &  $ x5$     &    $ x4$  \\
(viii) & $\frac{1}{2}+x2$    &  $\frac{1}{2}+x1$     &   $x3$     &  $ x4$     &    $-x5$  \\
\hline
\end{tabular} 
\end{table}

\clearpage

\begin{table}
\caption{Comparison of selected interatomic distances in the NbO$_{6}$-octahedra for SBN61 (neutron and X-ray data) and SBN34
(neutron data only). O4 and O5 are interchanged in this study and in the X-ray study  of Woike et al. \cite{Woike:2003}.
All distances are in  [\AA].}
\label{tbl:interatomic-distances-comparison}
\begin{tabular}{lcccccc}  
\hline
sample  & \multicolumn{2}{c}{SBN61}     & \multicolumn{2}{c}{SBN61}                           & \multicolumn{2}{c}{SBN 34} \\
dataset & \multicolumn{2}{c}{1/neutron} & \multicolumn{2}{c}{Xray/Woike \cite{Woike:2003}} & \multicolumn{2}{c}{2/neutron} \\
\hline
 & dmin & dmax & dmin & dmax &dmin &dmax \\
 \hline
 Nb2-O4         & 1.721(14) & 2.029(15) & 1.79(3) & 1.98(3) & 1.746(19) & 2.010(19) \\
 Nb2-O4$^{iv}$  & 1.959(14) & 2.213(15) & 1.95(3) & 2.18(3) & 2.022(19) & 2.280(19) \\
\hline
\end{tabular} 
\end{table}

\end{document}